\documentstyle[twocolumn,mathptm]{mn}  
  
\newcommand{\gtapprox}{\ga} 
\newcommand{\ltapprox}{\la} 
  
\input{psfig.sty}  
  
\begin{document}  
  
\title{Giant cluster arcs as a constraint on the scattering 
cross-section of dark matter} 
 
\author[Meneghetti et al.]{Massimo 
  Meneghetti$^{1,2}$, Naoki Yoshida$^2$, Matthias Bartelmann$^2$, 
  \newauthor Lauro Moscardini$^1$, Volker Springel$^3$, Giuseppe 
  Tormen$^1$ and  Simon D. M. White$^2$\\ 
  $^1$Dipartimento di Astronomia, Universit\`a di Padova, 
  vicolo dell'Osservatorio 2, I--35122 Padova, Italy\\ 
  $^2$Max-Planck-Institut f\"ur Astrophysik, Karl-Schwarzschild-Strasse 1, 
  D--85748 Garching, Germany \\ 
  $^3$Harvard-Smithsonian Center for Astrophysics, 60 Garden Street, 
  Cambridge, MA 02138, USA} 
  
\date{Accepted 2001 ???? ???; Received 2000 ???? ???;  
in original form 2000 ???? ??}  
  
\maketitle  
  
\begin{abstract}  
  
We carry out ray tracing through five high resolution simulations 
of a galaxy cluster to study how its ability to produce giant  
gravitationally lensed arcs is influenced by the collision 
cross-section of its dark matter. In three cases typical dark matter  
particles in the cluster core undergo between 1 and 100 collisions  
per Hubble time; two more explore the long (``collisionless'') and short 
(``fluid'') mean free path limits. We study the size and shape 
distributions of arcs and compute the cross-section for producing 
``extreme'' arcs of various sizes. Even a few collisions per particle 
modify the core structure enough to destroy the cluster's 
ability to produce long, thin arcs. For larger collision frequencies 
the cluster must be scaled up to unrealistically large masses before 
it regains the ability to produce giant arcs. None of our  
models with self-interacting dark matter (except the ``fluid'' limit) 
is able to produce radial arcs; even the case with the smallest  
scattering cross-section must be scaled to the upper limit of 
observed cluster masses before it produces radial arcs. Apparently 
the elastic collision cross-section of dark matter in clusters must be  
very small, below 0.1 cm$^2$g$^{-1}$, to be compatible with the 
observed ability of clusters to produce both radial arcs and giant 
arcs. 
  
\end{abstract}  
  
\begin{keywords}  
dark matter -- gravitational lensing -- cosmology: theory -- galaxies:  
clusters  
\end{keywords}  
  
\section{Introduction}  
  
Several recent observations indicate a potential problem with the Cold 
Dark Matter (CDM) scenario of structure formation in that they seem to 
contradict the numerically simulated density structure of CDM haloes. A 
first difficulty is posed by published rotation curves of dwarf galaxies, 
which often rise linearly from the centre to radii greater 
than $\sim1\,$kpc. Such a rise implies a constant-density core for the 
host dark-matter halo \cite{moore94,FloP94}, and is in conflict with 
the central density cusps of the haloes found in $N$-body 
simulations, which are usually fitted by a double power-law profile 
with central logarithmic slope $\simeq1\to1.5$ (Navarro, Frenk \& 
White 1997, hereafter NFW; Moore et al.~1999b). A second problem is 
related to the large number of low-mass subclumps found orbiting 
within simulated galaxy-size haloes; with a naive scaling to the 
observed properties of dwarf galaxies within the Local Group, the 
abundance of these subclumps appears too high by a factor of 
10 to 50 \cite{moore99,klypin99}. 
  
A possible way to modify the CDM scenario to be compatible with these 
observations has recently been suggested by Spergel  
\& Steinhardt \shortcite{spergel00}. They propose a model where the dark  
matter is self-interacting but dissipationless. They argue that if 
the dark matter particles have a sufficiently large cross-section for 
elastic scattering, the strength of the central density 
concentration and the abundance of orbiting subclumps can both be 
reduced. Collisions between dark matter particles would destroy the 
central cusps and lead to a gradual 
evaporation of small subclumps. 
  
Self-interacting dark matter would also affect the structure of galaxy 
clusters, making them more nearly spherical and reducing the number of 
subclumps. These effects have been confirmed by several recent studies 
in which the halo properties in cosmological simulations of 
collisionless cold dark matter and self-interacting dark matter have 
been compared. For instance, Yoshida et al. \shortcite{yoshida00b} and 
Dav\'e et al. \shortcite{dave00} found that intermediate scattering 
cross sections for dark matter particles produce haloes which are less 
centrally concentrated than in a collisionless model and have smoother 
and more spherical cores. Similar results have also been obtained by 
Burkert \shortcite{burkert00}, Firmani et al.~(2000,2001) and Kochanek 
\& White \shortcite{kochanek00}, although they find different time 
scales for the onset of core collapse, which transforms intermediate 
flat cores into cores with density profiles $\propto r^{-2}$. 
  
It is natural to ask how self-interacting dark matter would 
affect the strong gravitational lensing properties of galaxy 
clusters. Strong lensing is a very powerful tool to probe the 
distribution of dark matter in cluster cores. From the location 
of multiple images of background sources lensed by a cluster one can 
obtain a detailed map of the mass distribution in the lens 
\cite{tyson98}. The morphology of long arcs observed in the 
cluster fields yields important information on the core density 
\cite{bartelmann98}. Finally, the location of radial arcs can put 
strong constraints on the size and compactness of the lens cores 
(e.g.~Narayan \& Bartelmann 1997). 
  
The goal of this paper is to evaluate whether self-interacting 
dark-matter models are compatible with the observed ability of galaxy 
clusters to produce strong lensing events. Some recent work has 
already addressed this issue. Miralda-Escud\'e \shortcite{miralda00} 
argued that the shallow cores produced by self-interacting dark matter 
could not agree with current observations of lensing by 
clusters. Wyithe, Turner \& Spergel \shortcite{wyithe00} studied 
simplified analytic models of self-interacting dark matter haloes and 
found large collision cross-sections to be incompatible with multiple 
imaging by clusters. In this paper, we address the issue by using 
ray-tracing through high resolution simulations to compute lensing 
cross sections for arcs of various types.  We study a set of 
simulations designed to isolate the effect of collisions between dark 
matter particles on the structure of an individual cluster. 
 
The plan of our paper is as follows. 
In Section~\ref{method} we describe the cluster models used in the 
simulations and present the numerical method adopted to compute the 
lensing properties. We present and discuss the results of our analysis 
in Section~\ref{results}. Finally, our conclusions are drawn in 
Section~\ref{conclusion}. 
  
\section{Numerical simulations}  
\label{method}  
  
\subsection{Cluster models}  
  
The cluster models used in this study are those described in Yoshida 
et al.~(2000a,b). They consist of a set of five different 
resimulations of the same cluster-sized halo, which is the second most 
massive object in the GIF-$\Lambda$CDM simulation of Kauffmann et 
al. \shortcite{kauffmann99}. All simulations were performed with the 
GADGET code (Springel, Yoshida \& White 2000), in a flat model 
universe, with a matter density parameter $\Omega_{\rm m}=0.3$ and a 
cosmological constant of $\Omega_\Lambda=0.7$. The Hubble constant is 
assumed to be $H_0=70\,{\rm km\,s^{-1}\,Mpc^{-1}}$. The CDM power 
spectrum was normalised so that the {\em rms\/} matter density 
fluctuations in spheres of radius $r=8\,h^{-1}\,{\rm Mpc}$ is 
$\sigma_8=0.9$ ($h$ is the value of $H_0$ in units of 100 km s$^{-1}$ 
Mpc$^{-1}$). 

The first cluster of the series (hereafter called the S1 model) is 
just a resimulation at higher resolution of the original 
GIF-$\Lambda$CDM cluster, and was produced assuming standard 
collisionless dark matter. Three other simulations (hereafter called 
W-models) were carried out introducing elastic scattering between CDM 
particles. This was accomplished by the Monte-Carlo method proposed by 
Burkert \shortcite{burkert00}, slightly modified by considering the 
pairwise velocity difference in evaluating the scattering probability 
instead of the one-point velocity dispersion.  The three simulations 
differ in the value of the scattering cross section per unit mass 
$\sigma_\star$: $0.1\,{\rm cm^2\,g^{-1}}$, $1.0\,{\rm cm^2\,g^{-1}}$ 
and $10.0\,{\rm cm^2\,g^{-1}}$ were chosen. Hereafter, we will refer 
to these models as S1Wa, S1Wb and S1Wc, respectively, following the 
notation of Yoshida et al. \shortcite{yoshida00b}.  Finally, Yoshida 
et al. \shortcite{yoshida00a} also considered the extreme case in 
which dark matter particles are as strongly interacting as in the 
``fluid'' limit (see also Moore et al.~2000). In this last simulation, 
the time evolution, starting from the same initial conditions as in 
the collisionless original simulation, was followed solving the fluid 
equations with the SPH technique. In the following, this model will be 
referred to as S1F. 
 
The resolution achieved in these simulations is quite high. Most of 
them employ $0.5\times10^6$ particles in the central, high-resolution 
region, where the particle mass is  
$m_{\rm p}=0.68\times10^{10}\,h^{-1}\,M_\odot$. The gravitational 
softening length is set to $20\,h^{-1}\,{\rm kpc}$. All haloes have a 
similar final virial mass of $7.4\times10^{14}\,h^{-1}\,M_\odot$. 
 
For our lensing analysis, we picked the simulation snapshots at 
redshift $z=0.278$ because this is close to the most efficient 
redshift for strong gravitational lensing (e.g.~Bartelmann et 
al. 1998). 
 
\begin{figure} 
{\centering\leavevmode 
  \psfig{file=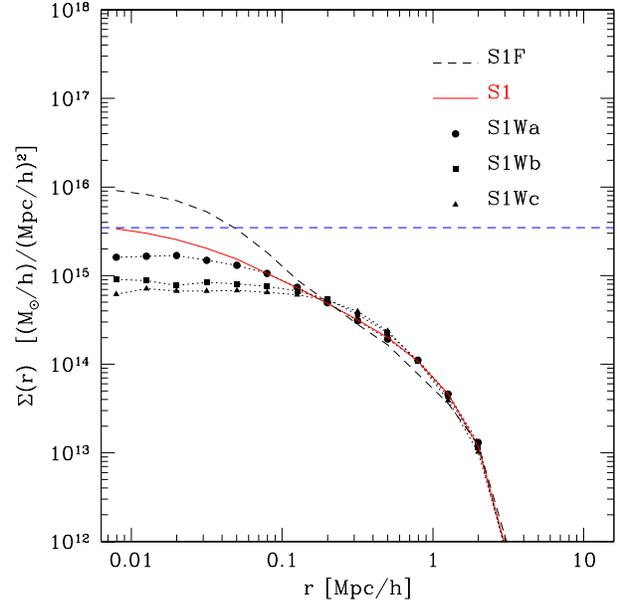,width=\hsize} 
} 
\caption{Surface density profiles of the five halo models at redshift 
$z=0.278$. The horizontal dashed line indicates the critical surface 
density for a lens at the same redshift and sources at redshift 
$z=2$. Note that large arcs are formed at surface densities below the 
critical value because of the shear.} 
\label{profiles} 
\end{figure} 
 
The two-dimensional density profiles of the five cluster models at 
this redshift are shown in Fig.~\ref{profiles}. The general trends in 
the profiles follow our expectations. As the scattering cross section 
of the dark matter particles increases, the cluster density profile in 
the inner region of the halo flattens. Note that the central density 
for the model S1Wb is smaller by more than one order of magnitude than 
that of the model S1. 
 
The halo shape also depends on $\sigma_\star$. The principal axis 
ratios of the five models, determined from the inertial tensors of the 
matter at densities exceeding 100 times the critical value, are listed 
in the third column of Table~\ref{clusters}. The values clearly show 
that the haloes are more spherically symmetric when the frequency of 
collisions between the dark matter particles increases. In the fourth 
column of Table~\ref{clusters}, we also report the core radius $r_{\rm 
c}$ of the three haloes with self-interacting dark matter,  
which we define as the distance from the cluster centre where the 
surface density drops below half its central value. For models S1Wb 
and S1Wc we typically find core radii larger than $200\,h^{-1}\,{\rm 
kpc}$. 
  
\begin{table} 
\caption{The main properties of the cluster haloes at redshift 
$z=0.278$. Column 1: halo model; column 2: scattering cross section 
per unit mass $\sigma_\star$; column 3: axial ratios; column 4: core 
radius $r_c$, defined as the clustercentric distance where the 
surface density falls below half its central value.} 
\begin{center} 
\begin{tabular}{lrcr} 
\hline\hline 
Model & $\sigma_\star$ & $a:b:c$ & $r_{\rm c}$ \\ 
& [${\rm cm^2\,g^{-1}}$] & & [$h^{-1}\,{\rm kpc}$] \\ 
\hline\hline 
S1    &  -   & 1:0.58:0.54 &  40 \\ 
S1Wa  &  0.1 & 1:0.63:0.60 & 110 \\ 
S1Wb  &  1.0 & 1:0.67:0.65 & 245 \\ 
S1Wc  & 10.0 & 1:0.86:0.85 & 400 \\ 
S1F   &  -   & 1:0.95:0.81 &  35 \\ 
\hline\hline 
\end{tabular} 
\end{center} 
\label{clusters} 
\end{table} 
 
\subsection{Lensing properties of the clusters} 
\label{lp} 
 
Our method to study the lensing properties of galaxy clusters has been 
described in detail in Meneghetti et al. \shortcite{meneghetti00}. 
Compared to that paper, only few parameters are changed here; 
therefore, we only give a brief description here and refer the reader 
to the original paper for further details. For a general introduction 
to the theory of gravitational lensing, see, e.g.~Schneider, Ehlers \& 
Falco \shortcite{schneider92}, Narayan \& Bartelmann 
\shortcite{narayan97} and references therein. 
 
Starting from a cluster simulation, we construct cluster lenses as 
follows. We centre the cluster in a cube of $3\,h^{-1}\,{\rm Mpc}$ 
side length and obtain the three-dimensional density field $\rho$ by 
interpolating the mass density within a regular grid of $256^3$ cells, 
using the {\em Triangular Shape Cloud\/} method (TSC; see Hockney \& 
Eastwood 1988). We then produce three different surface-density fields 
$\Sigma$ per cluster by projecting $\rho$ along each of the three 
coordinate axes. This yields three lens planes per cluster, which we 
consider independent cluster models for the present purpose.  As long 
as the TSC smoothing kernel is sampled by sufficiently many particles, 
smoothing and projecting can be interchanged, so that smoothing the 
three-dimensional density prior to projecting it is equivalent to 
projecting first and then smoothing the two-dimensional density. We 
checked and confirmed that changing the grid resolution up by a factor 
of two yields surface-mass densities with indistinguishable lensing 
properties. 
 
The surface density fields are then scaled by the critical surface 
mass density for lensing, which depends on the cosmological parameters 
and on the lens and source redshifts. We recall that the redshift for 
the lensing cluster is $z_{\rm L}=0.278$, and we put all sources to 
$z_{\rm S}=2$. This choice is motivated by the fact that galaxy 
clusters at redshifts between $0.2$ and $0.4$ are most efficient 
strong gravitational lenses for sources at redshifts $z\gtapprox1$. 
Although real sources are distributed in redshift, putting all of them 
at a single redshift is admissible because the critical surface 
density changes very little with source redshift if the lens redshift 
is substantially smaller, as is the case here. 
 
Scaling with the critical surface density yields three two-dimensional 
convergence fields $\kappa$ for each cluster. We then propagate a 
bundle of $2048\times2048$ light rays through the central quarter of 
each of these fields. Their deflection angles, directly obtained from 
the convergence, are used to compute the shear $\gamma$ and the 
elements of the Jacobian matrix $A$, which describes the local 
properties of the lens mapping. It is symmetric and can thus be 
diagonalised. Its two eigenvalues are 
\begin{equation} 
  \lambda_{\rm t}=1-\kappa-\gamma 
  \quad\mbox{and}\quad 
  \lambda_{\rm r}=1-\kappa+\gamma\;. 
\end{equation} 
 
Radial and tangential critical lines are located where the conditions 
\begin{equation} 
  \lambda_{\rm t}=0 \quad\mbox{and}\quad 
  \lambda_{\rm r}=0 
\end{equation} 
are satisfied, respectively. The corresponding caustics in the source 
plane, close to which sources are imaged as large arcs, are obtained 
applying the lens equation to the critical curves. All lens properties 
are computed on grids with an angular resolution of $0.19''$ on the 
lens plane, so that lensed images are properly resolved. 
 
The sources are initially distributed on a regular grid in the source 
plane. Their spatial density is iteratively increased near caustic 
curves. Placing a larger number of sources where the lens strength is 
highest increases the probability of producing long arcs and thus the 
numerical efficiency of the method. In the following statistical 
analysis, it is then necessary to compensate for this artificial 
increase by assigning to each image a statistical weight proportional 
to the inverse of the resolution of the grid on which its source was 
placed. 
 
Sources are assumed to be elliptical, with axis ratios randomly drawn 
from the interval $[0.5,1]$, and area equal to that of a circle of 
$1''$ diameter. We checked that changing the average source size does 
not affect our final results. 
 
For the classification of images we follow the technique introduced by 
Bartelmann \& Weiss \shortcite{bartelmann94}, which was also used in 
Meneghetti et al. \shortcite{meneghetti00}. 
 
\section{Results} 
\label{results} 
 
\begin{figure*} 
{\centering\leavevmode 
  \psfig{file=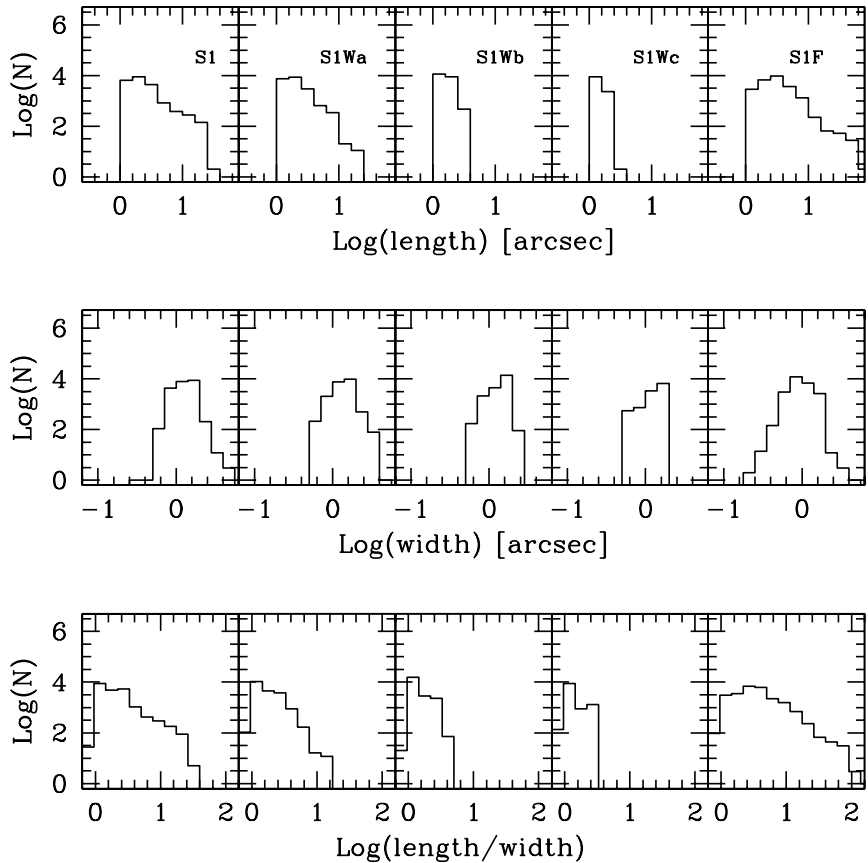,width=0.7\hsize} 
} 
\caption{Histograms of the image lengths (top panels), widths (middle 
panels) and length-to-width ratios (bottom panels). Images, whose area 
is smaller than two times the area of the unlensed source, are not 
considered here. Each column refers to a different model; from left to 
right: the pure collisionless case S1, the three models S1Wa, S1Wb and 
S1Wc, and the ``fluid'' limit model S1F.} 
\label{histos} 
\end{figure*} 
 
\subsection{Distributions of image properties} 
 
In this section, we present and compare the statistical properties of
the images of background sources lensed by the models of our cluster
sample. For each image, we compute the length $l$ (defined as the
maximum length of the circular segment passing through the image), the
width $w$ (found by fitting the image with several geometrical
figures; for further details, see Meneghetti et
al. 2000), and the length-to-width ratio
$l/w$. For the following analysis, we combine the results of two sets
of lensing simulations, in which we used the same lenses but changed
the distribution of the background sources by drawing different
orientations and axial ratios. With this procedure we confirm the
robustness of the results and improve the statistics.
Searching for large arcs, we can neglect the images of unlensed or 
even weakly lensed sources. Therefore, we can exclude those images 
from the analysis whose area is smaller than two times the area of 
their unlensed sources, which also avoids uncertainties in the 
automatic arc classification and parameter estimation. The number of 
images in the final analysed samples ranges between 11,252 for model 
S1Wc and 24,695 for model S1F. 
 
\begin{figure} 
{\centering\leavevmode 
  \psfig{file=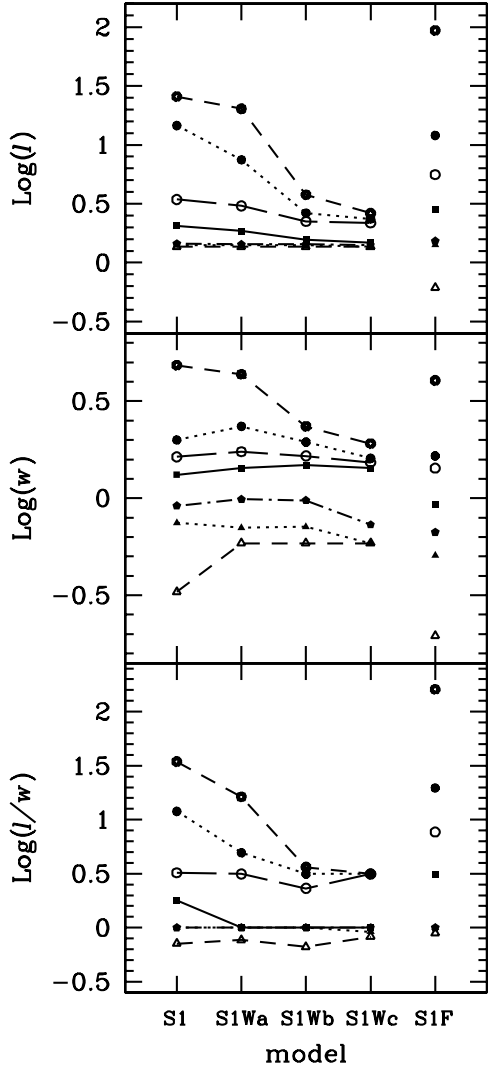,width=\hsize} 
} 
\caption{Percentiles of the arc length, width, and length-to-width 
ratio distributions are shown at the levels of $0\%$ (open triangles 
and dashed lines), $1\%$ (filled triangles and dotted lines), $10\%$ 
(filled pentagons and dotted-dashed lines), $50\%$ (filled squares and 
solid lines), $90\%$ (open circles and long-dashed lines), $99\%$ 
(filled circles and dotted lines) and $100\%$ (heavy open circles and 
dashed lines). The figure shows the lengths (or widths or 
length-to-width ratios) below which fall 0, 1, 10, 50, 90, 99, and 100 
per cent of the simulated arcs for the different cluster models. For 
instance, it can be read off the top panel that all arcs produced by 
model S1 are shorter than $25''$.} 
\label{moments} 
\end{figure}

We plot the distributions of arc properties in Fig.~\ref{histos}. The 
histograms show the number of arcs observed in the field of the five 
cluster models as a function of their length (top panels), width 
(middle panels) and length-to-width ratio (bottom panels). Each column 
refers to a different model. From left to right: the pure 
collisionless case S1, the three models S1Wa, S1Wb and S1Wc (with 
cross section of $0.1$, $1.0$, and $10.0\,{\rm cm^2\,g^{-1}}$, 
respectively), and the ``fluid'' limit model S1F. To facilitate the 
comparison, we show in Fig.~\ref{moments} the 0, 1, 10, 50, 90, 99, 
and 100 percentiles of the distributions plotted in Fig.~\ref{histos}. 
 
Regarding arc lengths, we find that model S1 produces a higher number 
of long arcs than the collisional W-models. Moreover, as shown in the 
top left panel of Fig.~\ref{moments}, the highest percentiles 
(i.e.~the largest arc lengths in the sample), change significantly 
going from S1 and S1Wa to S1Wb and S1Wc. This is easily understood in 
terms of the different density profiles of the models in the inner 
regions of the cluster. S1 has a steeper profile, and so it reaches a 
higher central surface density than models S1Wa, S1Wb and S1Wc. The 
tangential critical curve is located where $\lambda_{\rm t}=0$, 
i.e.~where $\kappa=1-\gamma$, and $\kappa$ is just proportional to the 
lens surface density. Therefore, the curve shrinks as the surface 
density profile flattens. Consequently, the probability of producing 
long arcs is highest for model S1, which has the longest tangential 
critical curve. 
 
Models S1Wb and S1Wc do not produce arcs longer that $\sim3''$, hence 
these models are unable to strongly distort background images. Their 
central surface density is $\ltapprox55\%$ and $\ltapprox50\%$ of the 
critical value, respectively, and the shear is reduced by the 
reduction of subhaloes and the increased axial symmetry caused by 
dark-matter interaction. Consequently, the combination of convergence 
and shear never succeeds to produce critical lines, regardless of the 
projection considered. At the opposite end, the fluid limit case S1F 
yields a number of long arcs larger than S1 because the extremely high 
density in the centre of this cluster pushes the tangential critical 
curve outward and extends it. 
 
The distribution of image widths is also sensitive to the size of the 
interaction cross section. The zero-percentile (giving the width of 
the thinnest arcs in the sample) moves to higher widths as 
$\sigma_\star$ increases, indicating that S1 produces thinner arcs 
than the W-models. As demonstrated by Kovner \shortcite{kovner89} and 
Hammer \shortcite{hammer91}, the radial extension of a tangential arc 
is magnified by a factor of order $\mu_{\rm r}\sim[2(1-\kappa)]^{-1}$ 
relative to its original size. Therefore, if $\kappa\ltapprox0.5$ at 
the tangential critical line, tangential arcs are demagnified in 
width. This condition can be achieved only if the central density of 
the lens is particularly high with a steep surface-density profile, as 
is the case for model S1F, or if the lens has an aspherical mass 
distribution or pronounced substructures 
\cite{bartelmann94,bartelmann95}. In fact, these lens features both 
contribute to pushing the critical curves outward, where $\kappa$ is 
lower. As the steepness of the density profile and the predominance of 
substructures both decrease going from model S1 to the W-models, the 
observed increase of the zero-percentile shown in the second panel of 
Fig.~\ref{moments} is consistent with our expectations. The lack of 
substructures in model S1F is compensated by the very dense core, 
providing this cluster with the capability of producing very thin 
arcs. 
 
From the middle panels in Fig.~\ref{histos} and the middle panel of 
Fig.~\ref{moments} (100-percentile), it can also be seen that the 
number of thick arcs decreases as the interaction cross section 
$\sigma_\star$ grows. The width distribution for S1Wb and S1Wc 
reflects the fact that these two models produce only weak distortions 
on the images of the background sources, that is, both image lengths 
and widths are only slightly changed with respect to their intrinsic 
values. On the other hand, the higher central surface density of the 
S1, S1Wa and S1F models makes them critical for strong lensing. 
 
Finally, and consistently with the previous discussion on arc lengths 
and widths, the distributions of the image length-to-width ratios show 
that, when the dark matter particles interact more frequently and the 
density profile becomes shallower, the number of arcs with large 
length-to-width ratio decreases. That is not the case for model S1F, 
where the density profile is even steeper than in model S1, causing 
the creation of many very long and thin arcs. 
  
\subsection{Lensing cross sections for long and thin arcs}  
  
We now compute absolute lensing cross section of the clusters for long 
and thin arcs. Arc cross sections for a property $Q$ are defined as 
the area in the source plane within which a source has to lie in order 
to be imaged as an arc with property $Q$. We consider here the cross 
sections for arcs with length-to width ratio larger than three 
different thresholds, namely: 
\begin{eqnarray} 
Q_1 &:& (l/w) > 5\nonumber\\ 
Q_2 &:& (l/w) > 7\nonumber\\ 
Q_3 &:& (l/w) >10\nonumber 
\end{eqnarray} 
In particular, the images which satisfy condition $Q_3$ are commonly 
called {\em giant\/} arcs \cite{wu93}. We are particularly interested 
in arcs with large length-to-width ratio because they have been 
observed in many galaxy clusters, and their statistics has been used 
in previous studies in an attempt to constrain cosmological parameters 
(Bartelmann et al.~1998 and references therein). 
  
\begin{table*}  
\caption{Lensing cross sections of the five halo models. Column 1: 
halo model; column 2: cross section $\sigma_{Q_1}$ for arcs with 
$l/w>5$; column 3: cross section $\sigma_{Q_2}$ for arcs with $l/w>7$; 
column 4: cross section $\sigma_{Q_3}$ for arcs with $l/w>10$; column 
5: cross section $\sigma_{Q_R}$ for radial arcs. All cross sections, 
given in units of $({\rm Mpc}/h)^2$, are averaged over the three 
projections of each cluster and two source-plane realisations for each 
projection.} 
\begin{tabular}{lcccc}  
\hline\hline  
Model &  
$\sigma_{Q_1}$ & $\sigma_{Q_2}$ &  
$\sigma_{Q_3}$ & $\sigma_{Q_R}$ \\  
\hline\hline  
 S1    & $6.93\times10^{-3}$ & $4.82\times10^{-3}$ &  
         $2.20\times10^{-3}$ & $8.13\times10^{-4}$ \\  
 S1Wa  & $1.46\times10^{-3}$ & $2.37\times10^{-4}$ &  
         $1.26\times10^{-4}$ & $0$\\  
 S1Wb  & $0$ & $0$ & $0$ & $0$\\  
 S1Wc  & $0$ & $0$ & $0$ & $0$\\  
 S1F   & $3.83\times10^{-2}$ & $2.36\times10^{-2}$ &  
         $1.14\times10^{-2}$ & $1.02\times10^{-2}$ \\  
\hline\hline  
\end{tabular} 
\label{crosssec}  
\end{table*}  
  
Each cluster model provides six estimates of its strong-lensing cross 
sections, two for each projection since we performed simulations with 
two source distributions. We average all measurements in order to 
obtain the mean cross sections for each cluster model. The lensing 
cross sections of the five models are listed in 
Tab.~\ref{crosssec}. Models S1Wb and S1Wc have vanishing cross 
sections even for arcs satisfying condition $Q_1$, given that, as 
shown in Fig.~\ref{moments}, they do not produce arcs with 
length-to-width ratio larger than $\sim3.5$. 
  
We find that model S1 has cross sections larger than model S1Wa. In 
particular, the cross sections $\sigma_{Q_2}$ and $\sigma_{Q_3}$ of 
the first model are more than one order of magnitude larger than those 
of the second. This feature suggests that a galaxy cluster consisting 
even of very weakly interacting dark matter should produce at least 
ten times fewer giant arcs than one made of collisionless 
particles. Compared to the other models, S1F has very large cross 
sections, in agreement with the results shown earlier.

As a consequence of the lower curvature in the effective lensing
potential, arcs produced by flatter profiles tend to be more strongly
magnified. However, the corresponding change in the magnification bias
is expected to have a very small effect on the previous results for
two reasons. First, such arcs would then be very thick, while the
observed ones are generally thin and sometimes unresolved in the
radial direction even when observed from space. Second, the
magnification bias depends on the colour selection, because the
number-count slope does, and unless only very blue or very red objects
are selected, the bias is negligible for arcs.
  
A comparison of the cross sections listed in Tab.~\ref{crosssec} for 
the collisionless simulation (S1) shows that the cluster almost 
exactly reproduces the average cross section found in the cluster 
sample used by Bartelmann et al.~(1998) for a lens redshift of 
$0.28$. In addition, we checked whether different realisations of the 
same cluster had comparable large-arc cross sections. To do so, we 
took the cluster mass distributions at different output redshifts 
($z=0.20$, $0.13$ and $0.06$) and computed their lensing properties 
when placed at a redshift of $0.28$. For the given cosmological 
parameters, the linear growth factor changes by less than $20\%$ 
between redshifts $0.28$ and $0.06$, so that the effect of cluster 
growth is unimportant. We find only a modest growth in the large-arc 
cross sections when using the cluster mass distributions at redshifts 
$0.20$ and $0.13$.  Between redshift $0.13$ and $0.06$, however, the 
cluster undergoes a merger event with a clump of 
$10^{13}\,h^{-1}M_\odot$ and subsequent relaxation, which 
increases its central density and therefore also the large-arc cross 
section. The effect of the merger on the lensing cross section is 
larger for the collisional than for the collisionless model. The 
infalling clump increases surface mass density and shear. Since the 
surface density profile of the collisional cluster is flatter than 
that of the collisionless cluster, this increase in lensing efficiency 
affects a larger region in the cluster centre. While the ratio between 
the large-arc cross sections between the S1 and S1Wa models is roughly 
an order of magnitude during quiescent periods, it is reduced to 
approximately five during the merger event. 
For similar reasons, it is important to mention that the presence of a
dominant central galaxy could affect our different models in different
ways, being more effective on top of the flatter models. As a
consequence, the differences between collisional and collisionless
cases could be reduced. A detailed study of this effect is in
progress.
 
Nonetheless, we can conclude (1) that the chosen cluster is typical 
for clusters in the $\Lambda$CDM cosmology, and (2) that the large-arc 
cross section for the mildly collisional model S1Wa falls 
substantially below that of the collisionless model S1 for different 
realisations of the same cluster. 
  
\subsection{Radial arcs}  

In this subsection, we analyse the capability of our models to produce 
radial arcs. Such arcs have been observed in several galaxy clusters. 
For instance, two radial arcs exist in the core of MS~0440 
\cite{gioia98}. From X-ray observations, the mass of this cluster 
within $\sim350\,h^{-1}\,{\rm kpc}$ was estimated to be 
$0.65\times10^{14}\,h^{-1}\,M_\odot$, which is similar to that of the 
cluster we analyse here. 

The appearance of radial arcs and their position within the lensing 
cluster depend on both the slope of the projected mass profile and on 
the central density of the lens. The surface density must surpass a 
critical value in order for the radial critical curve to appear where 
the eigenvalue $\lambda_{\rm r}=1-\kappa+\gamma$ of the Jacobian 
matrix is zero. Moreover, the steeper the density profile is, the 
closer to the centre the radial arcs tend to move 
\cite{williams99}. The observed location of radial arcs close to the 
cluster mass centres suggests that the core radii of the host clusters 
must be quite small \cite{mellier93,smail96,hammer97,gioia98}. Due to 
the different slopes of the density profiles of the cluster models, we 
thus expect to find quite large differences in their capacity to form 
radial arcs (see e.g. Molikawa \& Hattori 2000). 
 
\begin{figure} 
{\centering\leavevmode 
  \psfig{file=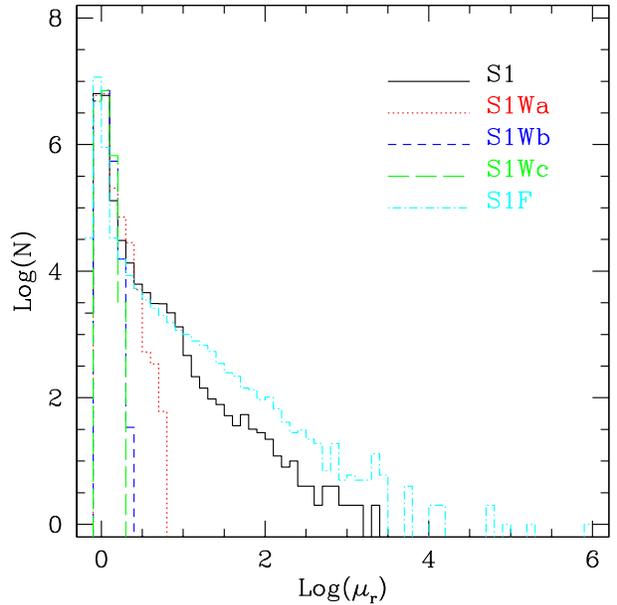,width=\hsize} 
} 
\caption{Histograms of the radial magnifications $\mu_{\rm r}$ at each 
point on the lens plane for the five cluster models: S1 (solid line), 
S1Wa (dotted line), S1Wb (dashed line), S1Wc (long-dashed line) and 
S1F (dot-dashed line).} 
\label{radhist} 
\end{figure}

The radial magnification $\mu_{\rm r}$ at each point of the lens plane 
is given by the inverse of the radial eigenvalue of the Jacobian 
lensing matrix, $\mu_{\rm r}=1/\lambda_{\rm r}$. From the convergence 
and the shear, computed as explained in Sect.~\ref{lp}, we obtain 
$\mu_{\rm r}$ at each of the $2048^2$ grid points of the lens 
plane. We show in Fig.~\ref{radhist} how these radial magnifications 
are distributed for the five cluster models. Each histogram combines 
contributions from all three projections of each cluster. As can be 
noted, values of $\mu_{\rm r}$ exceeding $10^3$ are reached only by 
models S1 and S1F, while models S1Wa, S1Wb and S1Wc can only produce 
weak radial distortions. 
 
Using the tangential and radial eigenvalues of the Jacobian matrix, we 
can also identify the radial arcs from the complete sample of 
distorted images, and this enables us to compute the lensing cross 
sections for radial arcs. For their automatic detection, we assumed 
that radial arcs must lie on the radial critical curve or very close 
to it. Therefore, we searched those images containing at least one 
pixel for which the radial eigenvalue is $\le0.15$ and the tangential 
eigenvalue is $\ge0.6$. These values were chosen in order to avoid 
misclassifications of radial arcs. 
 
The cross sections for radial arcs, $\sigma_{\rm R}$, averaged over 
the six simulations performed for each cluster, are listed in the 
fifth column of Table~\ref{crosssec}. We found that only models S1 and 
S1F are able to produce radial arcs, while the W-models are not. The 
results for model S1Wa are particularly interesting because this 
cluster seems to have a sufficiently high surface density to produce 
giant arcs, but not enough to radially distort the background 
galaxies. 
  
\subsection{Rescaling the clusters}  

Our results show that models S1Wb and S1Wc cannot produce long and 
thin arcs at all. Although the mass of these clusters is already 
fairly large, we have checked whether an even more massive cluster 
with density profile similar to that of S1Wb could efficiently form 
this kind of arcs. To this end, we rescaled the cluster mass by a 
variable factor $f>1$. 
 
The virial radius $R_{\rm vir}$ of a halo scales as $R_{\rm
vir}\propto M^{1/3}$, where $M$ is the virial mass of the halo. To
obtain a halo of mass $f\times M$ which is dynamically stable, we also
need to rescale distances by a factor $f^{1/3}$. This means that,
while the three-dimensional density $\rho$ remains fixed, the halo
surface density is enhanced by a factor $f^{1/3}$.  For this reason we
expect that, increasing its mass, cluster S1Wb will eventually become
able to produce strong lensing effects.
 
\begin{figure} 
{\centering\leavevmode 
  \psfig{file=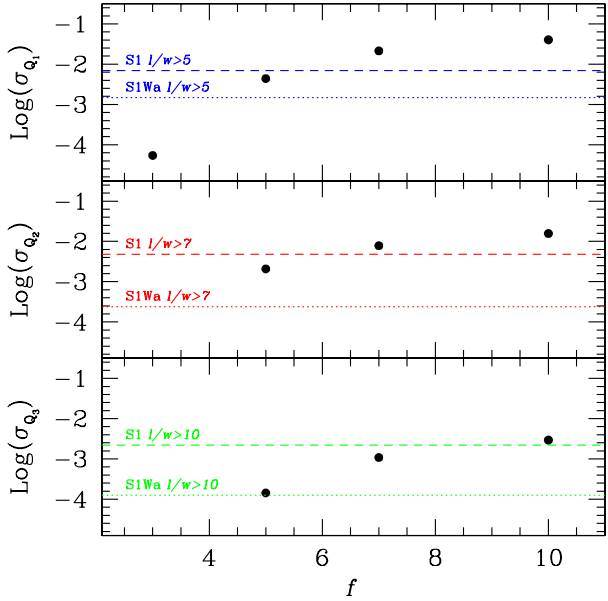,width=\hsize} 
} 
\caption{Lensing cross sections of cluster model S1Wb as a function of 
the rescaling factor $f$. The three panels show the cross sections for 
arcs with $l/w>5$ ($\sigma_{Q_1}$, top frame), $l/w>7$ 
($\sigma_{Q_2}$, middle frame) and $l/w>10$ ($\sigma_{Q_3}$, bottom 
frame). The dashed and dotted lines in each panel indicate the lensing 
cross sections of cluster models S1 and S1Wa, respectively. The cross 
sections are given in units of $(\mbox{Mpc}/h)^2$.} 
\label{rescaled} 
\end{figure}

We therefore perform a new set of lensing simulations, choosing mass 
scaling factors $f$ equal to 3, 5, 7, and 10. We plot in 
Fig.~\ref{rescaled} the cross sections of these rescaled clusters, as 
a function of the rescaling factor $f$. In this plot, upper, middle 
and lower panels refer to the lensing cross section for 
length-to-width ratios larger than 5, 7, and 10, respectively. The 
dashed and dotted horizontal lines correspond to the equivalent cross 
sections for models S1 and S1Wa. 
 
We can infer from this figure that a cluster with the same density 
profile as our model S1Wb must have a mass of at least $\sim5.5$ 
($\sim4.5$) times larger than the mass of our present cluster, in 
order to achieve the same strong lensing efficiency of model S1 
(S1Wa). 
 
In the same spirit, we also rescale the mass of cluster S1Wa, to 
determine when it starts producing radial arcs. We multiply the 
cluster mass by a factor of two and found that only one out of three 
projections of the rescaled cluster was efficient in forming radial 
arcs. However, the cross section of the rescaled cluster, $\sigma_{\rm 
R}=2.68\times10^{-5}$, is still one order of magnitude lower than that 
of model S1.    
Finally let us comment that our method to rescale the clusters is
approximate. In fact, rescaling the mass should in principle affect
the collision rate. Yoshida et al.~(2000b) estimated the number of
collisions between dark matter particles per Hubble time as $N_{\rm
coll}=\rho \sigma_{\star} V$, where $\rho$ is the three-dimensional
density and $V$ is the particle velocity. Since $V \propto M^{1/2}$,
the number of collisions would increase by a factor $f^{1/2}$ when the
cluster mass is rescaled by a factor $f$. Since our rescaling
technique keeps $N_{\rm coll}$ constant while the cluster mass is
increased by a factor $f$, the result practically corresponds to an
effective $\sigma_{\star}$ reduced by a factor $f^{1/2}$. This means
that we are even over-estimating the lensing efficiency of the
$f$-scaled clusters and our previous conclusions can be considered
conservative.

\section{Summary and Conclusions}  
\label{conclusion}  
  
Self-interacting dark matter was suggested as a solution to the 
problems (1) that dwarf galaxies are observed to have flat cores, 
while they should have density cusps according to $N$-body 
simulations, and (2) that simulations tend to produce more 
substructure in galaxies than there seems to be. In essence, the 
dark-matter self-interaction flattens density cusps, increases radial 
symmetry, and damps substructure; all of which are `desired' effects. 
  
However, it was shown before that asymmetries, steep density profiles, 
and existence of substructures are essential for the ability of galaxy 
clusters to produce pronounced strong-lensing features like large 
arcs. This is due to several reasons. First, lenses need to exceed the 
critical surface mass density in order to form critical curves, which 
are mandatory for large arcs. Second, asymmetries in lenses increase 
the gravitational tidal field, or shear, which helps to make them 
critical for multiple imaging. Asymmetric clusters can be critical at 
lower surface mass densities than symmetric clusters. Third, observed 
arcs are generally thin, i.e.~little magnified or even demagnified in 
the radial direction. For this to be the case, the surface mass 
density needs to be lower than about half the critical surface density 
at the location of the arcs. Together with the requirement that the 
surface density be supercritical in lens centres, this implies that 
density profiles have to be steep. 
  
In other words, the introduction of self-interaction between 
dark-matter particles is expected to have a pronounced, potentially 
`undesirable' impact on the ability of clusters to produce large arcs: 
it is exactly the desired effect of the self-interaction, namely to 
make clusters flatter, rounder, and smoother, that on the other hand 
threatens to destroy their strong-lensing capabilities. In this paper, 
we have investigated this effect on the second-most massive cluster in 
a cosmological $\Lambda$CDM simulation, and produced the following 
results: 
  
\begin{itemize}  
  
\item The introduction of self-interaction even with a small cross  
section of $\sigma_\star=0.1\,{\rm cm^2\,g^{-1}}$ reduces the cluster 
cross section for the production of large arcs by about an order of 
magnitude compared to the collisionless case. 
  
\item Upon further increase of the interaction cross section, the  
cluster becomes entirely uncritical and cannot produce arcs any more, 
despite the otherwise favourable conditions in terms of cluster mass 
and redshift. 
  
\item The cross section for radial arcs vanishes even for the smallest  
interaction cross section considered. 
  
\item The cluster with an interaction cross section of  
$\sigma_\star=1.0\,{\rm cm^2\,g^{-1}}$ would require about five times 
more mass to reach the cross section for large tangential arcs of the 
cluster with $\sigma_\star=0.1\,{\rm cm^2\,g^{-1}}$, and about $6-8$ 
times more mass to restore the cross section of the collisionless 
cluster. Due to the steepness of the cluster mass function, the 
spatial number density of such clusters at the redshift considered is 
more than six orders of magnitude smaller than that of the original 
cluster. 
  
\item The strong-lensing cross section of the cluster model  
investigated here is almost identical to the average cross section 
found in a sample of clusters simulated earlier in the $\Lambda$CDM 
cosmology (Bartelmann et al.~1998). Also, the cross sections obtained 
do not change significantly when the cluster model is taken at 
different output times and placed at the original lens redshift of 
$z=0.28$. 
  
\end{itemize}  
  
At a length-to-width ratio of $l/w\ge10$, the abundance of observed 
arcs brighter than $R=22.5$ is $\sim0.2-0.3$ per cluster with X-ray 
luminosity $\ge2\times10^{44}\,{\rm erg\,s^{-1}}$. Extrapolated to the 
whole sky, there should be $\sim1500-2300$ such arcs. Numerically 
simulated clusters in a (cluster-normalised) $\Lambda$CDM model fall 
short by about an order of magnitude of producing this number of arcs 
(Bartelmann et al.~1998). It may be possible that massive central 
cluster galaxies can increase the arc optical depth of $\Lambda$CDM 
clusters sufficiently to reconcile it with observations (Williams et 
al.~1999). However, a further reduction of the strong-lensing cross 
section by an order of magnitude due to even mild dark-matter 
interaction seems problematic in view of the observed arc abundance. 
 
Although the inability to produce radial arcs of the clusters with
interacting dark matter is a potentially important piece of
information, it is currently impossible to draw any firm conclusions
from comparisons with data. Radial arcs have so far been reported in
five galaxy clusters (MS~2137, Fort et al.~1992; A~370, Smail et
al.~1996; MS~0440, Gioia et al.~1998; AC~114, Natarajan et al.~1998;
A~383, Smith et al.~2000), all of which are located near bright
galaxies in the cluster centre. It is currently unclear what exactly
the influence of the galaxies on the occurrence of the radial arc is,
and how the presence of luminous galaxies near cluster centres
prevents the detection of more radial arcs. A statistical comparison
of predicted and observed radial-arc numbers therefore appears
premature. In principle, the ratio of the clustercentric distances of
radial and tangential arcs could be used to constrain the cluster
density profile, but in practice at least redshift estimates of the
arc sources would have to be available.
  
Of course, our simulations are insufficient for statistically sound 
statements on the total arc cross section of a given cluster 
population. However, given the massive impact of dark-matter 
interaction even with the smallest cross section considered, we 
conclude that the dark-matter self-interaction hypothesis may be in 
severe conflict with the abundance of large arcs unless the 
interaction cross section is very small. 
  
\section*{Acknowledgements}  

This work was partially supported by Italian MURST, CNR, CNAA and ASI,
and by the TMR european network ``The Formation and Evolution of
Galaxies'' under contract ERBFMRX-CT96-086. MM, LM and GT thank the
Max-Planck-Institut f\"ur Astrophysik, and MB the Dipartimento di
Astronomia di Universit\`a di Padova, for their hospitality during the
visits when this work was completed. We are grateful to the anonymous
referee for comments which allowed us to improve the presentation of
the results.

\end{document}